\documentclass[11pt,headings=small,numbers=enddot,twocolumn=on]{scrartcl}
\usepackage[T1]{fontenc}
\usepackage[utf8]{inputenc}
\usepackage{paralist}
\usepackage{rotating}
\usepackage{csquotes}
\usepackage[svgnames]{xcolor}
\usepackage{times}
\usepackage[config=mt-ptm]{microtype}

\usepackage[colorlinks=true, urlcolor=DarkSlateBlue,
citecolor=DarkSlateBlue, filecolor=DarkSlateBlue, plainpages=false,
pdfpagelabels, bookmarksnumbered]{hyperref}
\usepackage{etoolbox}

\usepackage[authordate, backend=biber, babel=hyphen, bibencoding=inputenc, strict, isbn=false, uniquename=false]{biblatex-chicago} % biblatex setup
\usepackage{graphicx}
\makeatletter
\def\maxwidth{\ifdim\Gin@nat@width>\linewidth\linewidth
\else\Gin@nat@width\fi}
\makeatother
\let\Oldincludegraphics\includegraphics
\renewcommand{\includegraphics}[1]{\Oldincludegraphics[width=\maxwidth]{#1}}

\title{Cultural and Demic Diffusion of First Farmers, Herders, and their
Innovations Across Eurasia}

\newif\ifblind
\blindtrue
\blindfalse

\newif\ifreview
\reviewtrue
\reviewfalse

\newif\ifcount
\countfalse
\ifcount
\reviewfalse
\KOMAoptions{twocolumn=off}
\else
\ifreview
\countfalse
\KOMAoptions{twocolumn=off}
\usepackage[margin=25mm,right=30mm,top=20mm,bottom=40mm]{geometry}
\usepackage{setspace}
\usepackage[nolists,noheads]{endfloat}
\usepackage{lineno} % add later in word
\linenumbers
\usepackage[none]{hyphenat}
\usepackage{ragged2e}
\RaggedRight

\else
\usepackage[left=17mm,right=17mm,top=15mm,bottom=20mm]{geometry}
\fi

\usepackage{scrpage2}
\ihead{}
\ihead{\normalfont\scshape Cultural and Demic Diffusion of First Farmers, Herders, and their
Innovations Across Eurasia}
\ihead{\normalfont\scshape Cultural and Demic Diffusion Across Eurasia}
\ifblind\else\ohead{\normalfont\scshape  Carsten Lemmen          }\fi
\ofoot[\itshape Page~\thepage]{\itshape Page~\thepage}
\ifoot{\normalfont\rmfamily\footnotesize   
Version \today
, submitted to \textit{Documenta Praehistorica}
}
\fi

\addtokomafont{subsection}{\rmfamily\normalsize\itshape}
\addtokomafont{section}{\rmfamily\normalsize\bfseries}

\author{\ifreview
\Large Carsten Lemmen\\ 
\parbox{\hsize}{\normalsize
}
\else
\Large Carsten Lemmen\\ 
\parbox{\hsize}{}
\fi
\newline
\ifreview
\Large \\ 
\parbox{\hsize}{\normalsize
{Affiliation: C.L. Science Consult, Lüneburg, Germany and Helmholtz-Zentrum
Geesthacht, Germany}
}
\else
\Large \\ 
\parbox{\hsize}{{\itshape\small, C.L. Science Consult, Lüneburg, Germany and Helmholtz-Zentrum
Geesthacht, Germany}
}
\fi
\newline
\ifreview
\Large \\ 
\parbox{\hsize}{\normalsize
{Address: Lauensteinstraße 30, 21339 Lüneburg, Germany, Fax +49 4131 927 102-9}%
}
\else
\Large \\ 
\parbox{\hsize}{{\itshape\small, Lauensteinstraße 30, 21339 Lüneburg, Germany, Fax +49 4131 927 102-9}%
}
\fi
\newline
\ifreview
\Large \\ 
\parbox{\hsize}{\normalsize
{Email: \href{mailto:science@carsten-lemmen.de}{science@carsten-lemmen.de}}
}
\else
\Large \\ 
\parbox{\hsize}{{, \small\href{mailto:science@carsten-lemmen.de}{science@carsten-lemmen.de}}
}
\fi
\newline
}

\newcommand{\degree}{\ensuremath{^\circ}}

\date{}

\begin{document}  

\ifreview\else\begin{onecolumn}\fi

\maketitle
\ifreview\newpage\fi

%\begin{abstract}
{
\noindent \textbf{Abstract.} Was the spread of agropastoralism from the Eurasian founder regions
dominated by demic or by cultural diffusion? This study employs a
mathematical model of regional sociocultural development that includes
different diffusion processes, local innovation and societal adaptation.
Simulations hindcast the emergence and expansion of agropastoral life
style in 294 regions of Eurasia and North Africa. Different scenarios
for demic and diffusive exchange processes between adjacent regions are
contrasted and the spatiotemporal pattern of diffusive events is
evaluated. This study supports from a modeling perspective the
hypothesis that there is no simple or exclusive demic or cultural
diffusion, but that in most regions of Eurasia a combination of demic
and cultural processes were important. Furthermore, we demonstrate the
strong spatial and temporal variability in the balance of spread
processes. Each region shows sometimes more demic, and at other times
more cultural diffusion. Only few, possibly environmentally marginal,
areas show a dominance of demic diffusion. This study affirms that
diffusion processes should be investigated in a diachronic fashion and
not from a time-integrated perspective.
}
%\end{abstract}

\ifreview\newpage\fi

\ifreview\else
\end{onecolumn}
\begin{twocolumn}
\setkeys{Gin}{width=1\textwidth} 	
\fi


\section{Introduction}\label{introduction}

The transition to agriculture and pastoralism, termed the ``Neolithic
revolution'' by Childe (1925) has fundamentally changed social systems
and the relationship of people and their environments. However
revolutionary - even termed \enquote{traumatic} (Rowley-Conwy 2004) -
this transition was locally, the more gradual it appears on the
continental scale, spanning almost 10000 years of human prehistory and
history (e.g, Barker 2006).

The spatial diffusion of the new agropastoral and animal husbandry
innovations, technologies, and life styles played a major part in the
abandonment of a foraging life style following local innovations in very
few places worldwide that are associated with the domestication of
plants and animals (Fuller et al. 2014). From these few founder regions,
the new domesticates, their cultivation knowledge and the idea of
farming and herding itself spread to all but the most secluded or
marginal regions of the world; not only these cultural traits spread,
but also people, who carried along their \enquote{hitchiking} traits
(Ackland et al. 2007).

The spatiotemporal pattern of dated Neolithic sites consequently
radiates outward from the founder regions. For different cultural and
individual traits, the apparent rates of spreading can be determined
(Edmonson 1961, Bocquet-Appel et al. (2012)), but it is unclear from the
spatiotemporal analysis of dated sites, what process dominated the
expansion (Lemmen, Gronenborn, and Wirtz 2011): Within a broad spectrum
of diffusion mechanisms that include, e.g., also leapfrog migration and
elite replacement (Zvelebil 1998) demic diffusion and cultural diffusion
represent two contrasting views that have received widespread attention
in the literature. The demic diffusion hypothesis suggests the
introduction of the new agropastoral technologies through movements of
people - migrations of any form; the cultural diffusion hypothesis
suggests a technology shift through indigenous adaptations and
inventions fostered by culture contact - information dispersal of any
form.

Demic diffusion, i.e.~the spread of agropastoralism by migration of
people has been put forward as one of the earliest hypotheses for
explaining the spatiotemporal pattern of Neolithic arrival dates in
Europe (Clark 1965); evidence for demic diffusions is accumulating with
modern mtDNA and Y-chromosomal analyses revealing matrilinear and
patrilinear relationships in space and time (Chikhi et al. 2002,
Deguilloux et al. (2012), Fu et al. (2012)) (although contrasting views
have been presented by Battaglia et al. (2008) and Haak et al. (2010)),
and with earlier linguistic work (Renfrew 1987).

Cultural diffusion is the spread of agropastoralism by information and
material transmission in the absence of migrations. As both maternal and
paternal genetic lines are continuous from the Founder regions into
Europe, approval for the cultural diffusion hypothesis depends on a
temporal mismatch between the expansion of traits and knowledge and the
expansion of people. Already Ammerman and Cavalli-Sforza (1973)
suggested that both demic and diffusive spread are active and that it is
the relative contribution of each that needs to be investigated rather
than deciding on either demic or cultural diffusion. Furthermore,
cultural diffusion theories have also been put forward as a reaction
against processual diffusionist views and emphasize the agency and
innovativity of local populations (Hodder 1990) (but refuted again by
e.g., Rowley-Conwy (2004)).

Mathematical models on the spread of agropastoralism have a long
tradition in Europe and can be traced back to Childe (1925)'s
observations on the spatio-temporal distribution gradient of ceramics
from Southeastern to Northwestern Europe. This pattern was replicated
from Neolithic radiocarbon dates by Clark (1965), and subsequently
mathematically formulated by Ammerman and Cavalli-Sforza (1973) as the
\enquote{wave of advance} model on which many subsequent formulations
have been built (Ackland et al. 2007,Galeta et al. (2011),Davison,
Dolukhanov, and Sarson (2009)).

A common feature of diffusion models is a concentric expansion from one
or multiple centers of supposed origins, with modifications introduced
to account for geographic bottlenecks, terrain, or rivers (Davison et
al. 2006,Patterson et al. (2010),Silva2014a). Fort (2012) and Fort
(2015) attempted to disentangle demic and cultural diffusion both from a
modeling as well as a data perspective. In a diffusion model, they found
that both demic and cultural diffusion are important, with demic
diffusion responsible for 60\% (vs.~40\% for cultural) of the spreading
process. Similarly, our own investigation (Lemmen, Gronenborn, and Wirtz
2011) concluded that a mixed model produces a pattern of Neolithization
best representing the data.

Much less numerical studies have been performed for Eurasian regions
outside Europe. The best investigated test case is probably South Asia
and the Indian subcontinent. For this region Ackland et al. (2007)
investigated the transition to agriculture as a diffusion process that
emanates from a single founder region in Southwest Asia; in contrast,
Patterson et al. (2010) reported on a simulation of the Neolithic
transition in India expanding from two centers, representing Chinese and
Harappan migration streams. Our own simulations for the Indian
subcontinent showed that the connection from the Indus region to the
Levante was only established after the transition to agropastoralism
(Lemmen and Khan 2012), consistent with the wheat/rice barrier
identified by (Barker 2006). The demic--cultural debate has not been
investigated for greater Eurasia yet.

In the current study, I demonstrate with numerical simulations how the
different assumptions about the diffusion process -interpreted as demic
diffusion and cultural diffusion or a mixture thereof- may have played
different roles in the spread of agropastoralism through Eurasia.
Emanating from founder regions in North and South China, Central Asia,
and the Levant about 9000 years ago, the entire continent (except
Northern Eurasia) transitions to agropastoral life styles by 3000~BC
drawing a complex picture of cultural and demic diffusion.

The goal of this study is to investigate qualitatively the spatial and
temporal predominance of either cultural or demic diffusion processes
within Eurasia, and to provide a novel visualization of the complexity
of the interplay between these processes at a continental scale.

\section{Methods}\label{methods}

I employ the Global Land Use and technological Evolution Simulator
(GLUES, Lemmen, Gronenborn, and Wirtz 2011)---a numerical model of
prehistoric innovation, demography, and subsistence economy---to
hindcast the regional transitions to agropastoralism and the diffusion
of people and innovations across Eurasia for the period 7500--3500~BC.

\begin{figure}[htbp]
\centering
\includegraphics{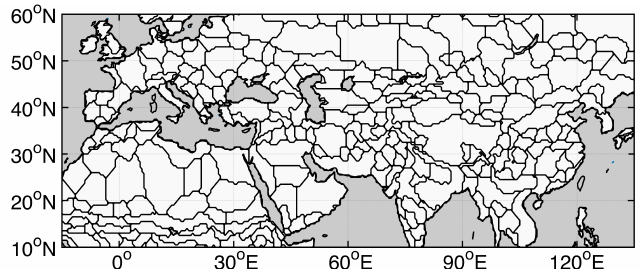}
\caption{Geographic setting of 294 Eurasian and North African simulation
regions in the Global Land Use and technological Evolution Simulator.
This is a subset of the full (global) simulation comprising 685 world
regions. \label{fig:regions}}
\end{figure}

The model operates on 294 (country-like) spatial units within the domain
-15 \degree{}E to 135 \degree{}E and 10 \degree{}N to 60 \degree{}N (\hyperref[fig:regions]{Figure 1}).
These regions represent ecozones that have been derived to represent
homogenous net primary productivity (NPP) clusters based on a 3000~BC 1\degree{}
x 1\degree{} palaeoproductivity estimate (Wirtz and Lemmen 2003); this estimate
was derived from a dynamic palaeovegetation simulation (Brovkin,
Ganopolski, and Svirezhev 1997) scaled down with the New et al. (2001)
climatology. By using NPP, many of the environmental factors taken into
account by other expansion or predictive modesl, such as altitude,
latitude, rainfall, or temperature (e.g, Silva2014b, Arikan2014).

Within each region, a trait-based adaptive model describes regional
societies with three characteristics: intrinsic innovations
(technology), extrinsic (economic diversity), and subsistence style
(Lemmen, Gronenborn, and Wirtz 2011). The evolution of these
characteristic traits is interdependent and drives the growth of a
regional population according to the gradient adaptive dynamics approach
formulated by Wirtz and Eckhardt (1996) for ecological systems. In his
approach, the rate of change of the mean of each characteristic trait is
calculated as the product of the trait's variability and its marginal
growth benefit, i.e.~the derivative of population growth rate with
respect to the trait, evaluated at the mean growth rate. In Wirtz and
Lemmen (2003), we adopted this mathematical approach for social systems;
as the approach is an aggregate formulation operating on the statistical
moments of traits and growth rate, it requires large populations, and
thus larger geographic areas. For further details on the trait-based
model formulation, see Lemmen, Gronenborn, and Wirtz (2011) (their
supplementary online material).

Exchange of characteristic traits and migration of people between
regions is formulated with a diffusion-like approach, i.e., the flow of
a quantity (technology, economic diversity, subsistence style) is
directed from a region with higher influence (i.e.~product of technology
and population) to a region with lesser influence. The speed of the
spread is proportional to the interregional difference of the respective
quantity and of influence, is proportional to the influential region's
technology, and proportional to common boundary length divided by
interregional distance. Migration is furthermore dependent on acceptable
living conditions (positive growth rate) in the influenced region.
Equations for interregional interchange are given in the appendix. The
size of the simulation regions (on average 300000 km$^2$) is
insufficient for detailed local analyses, but appropriate for
subcontinental and continental-scale simulations and necessary to allow
for parameter space exploration.

We performed three different simulations, one with mixed diffusion, one
with exclusively demic diffusion and one with exclusively cultural
diffusion (see appendix for the different formulations). The global
simulations (in total 685 regions) are started at 8500~BC, assuming
equal initial conditions for all societies in all regions; we use the
same set of parameters that have been used by Lemmen, Gronenborn, and
Wirtz (2011): for the three diffusion scenarios, we obtained the
diffusion coefficients by tuning each model to optimally represent the
European arrival dates. Simulations were performed with GLUES version
1.1.20a; this version can be obtained as free and open source from

Despite tuning all scenarios to the radiocarbon record used in Lemmen,
Gronenborn, and Wirtz (2011), the highest correlation could only be
obtained with the mixed (base) scenario. To disentangle cultural and
demic diffusion processes, we compared the demic and cultural diffusion
scenarios with each other after normalization with the mixed scenario.
Where the demic scenario predicted at least a 10\% greater share of
agropastoral life style, we diagnosed a predominantly demic diffusion.
Where the cultural scenario predicted a greater share, we diagnosed a
predominantly cultural diffusion. To estimate the overall influence of
demic versus cultural diffusion, we averaged for each region the
relative predominance of demic over cultural diffusion processes over
time.

\section{Results}\label{results}

The timing of the arrival of agropastoralism
(\hyperref[fig:timeslices]{Figure 2}) reveals its multicentric origin
and spatiotemporal expansion, including the typical radiation from
founder regions seen in all diffusive models.

\begin{figure}[htbp]
\centering
\includegraphics{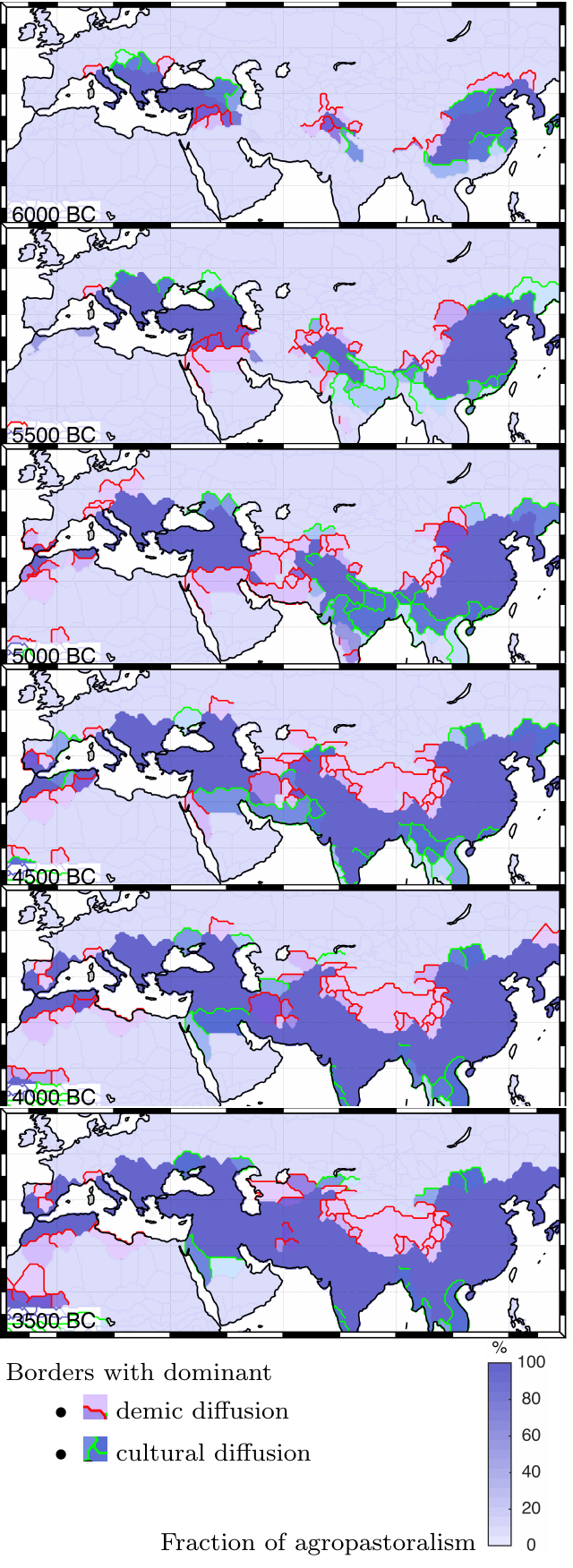}
\caption{Simulated transition to agriculture, 6000-3500~BC. The darker
the shading, the higher the fraction of agropastoralists in the
population. Red lines show regional borders with demic diffusion events,
green lines show regional borders with cultural diffusion
events.\label{fig:timeslices}}
\end{figure}

By 6600~BC, the transition to agropastoralism has occurred in five
founder regions: (1) Northern coastal China, (2) Southern tropical
inland China, (3) Northern Indus region, (4) West Anatolia and Greece,
and (5) Zagros mountains. At this time, emerging agropastoralism
connects the Chinese regions with each other
(\hyperref[fig:timeslices]{Figure 2}). By 6300~BC, agropastoralism is
the dominant life style in all founder regions; it has expanded west to
the Balkans and Italy, and east to Korea. A broad band of
agriculturalists is visible across China.

By 6100~BC, the Levante and Anatolian founder regions connect and expand
north and eastward, likewise the Chinese regions. The Indus regions
extends towards the Ganges. These emerging life styles consolidate in
the ensuing centuries. By 5500~BC, the western Eurasion center has
continued to expand in all directions, reaching around the Black Sea and
to the Caspian Sea. All of China has transitioned Emerging
agropastoralism connects the Indus to the Chinese region. By 5100~BC,
North African pastoralism emerges. There is now one large Asian
agropastoralist region, also with emergent transitions throughout India.

By 4700 the Western and Eastern Eurasian center connect. Agropastoralism
emerges in Southeast Asia and Western Europe. By 4000~BC, one large belt
of agropastoral life style connects the Mediterranean with West Asia,
South Asia, and East Asia.

Multiple, intermittent, and recurrent predominantly demic or cultural
diffusion processes are seen throughout the simulation for all regions.
For example, exchange processes around the Central Asian plateau are
dominated by demic diffusion at all times. At most times, North African
and Southwest European exchange processes are dominated by demic
diffusion. Cultural diffusion, on the other hand, is at all times
dominant within east and south China, and in Southeast Asia. It is at
most times dominant on the Indian subcontinent.

A more complex pattern of demic and cultural diffusion in space and time
is observed in Western Asia and Southeast Europe. Diffusion from the
Fertile Crescent is predominantly demic before 4900~BC, and cultural
thereafter. Just east of the Red Sea, it is demic until 4200~BC, and
cultural from 4000~BC. The expansion of Southeastern and Anatolian
agropastoralism northward is predominantly cultural at 5500~BC, and
predominantly demic 500 years later. At 5000~BC, it is demic west of the
Black sea and cultural east of the Black Sea. The, at 4500~BC, demic
processes again take over part of the eastern Black Sea northward
expansion.

\begin{figure}[htbp]
\centering
\includegraphics{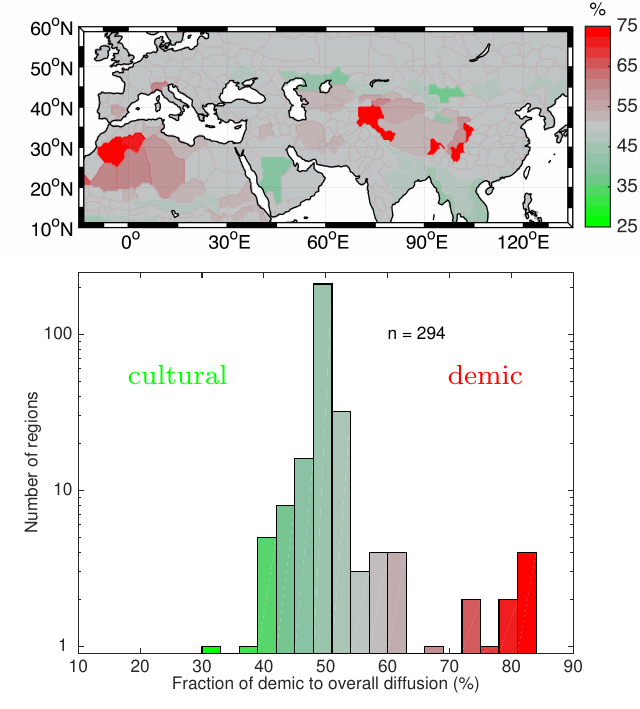}
\caption{Time integrated contribution of predominantly demic (red) and
cultural (green) diffusion represented geographically (top panel) and as
a histogram (bottom panel). For most regions, no predominance (grey) of
either mechanism is found. \label{fig:demiccultural}}
\end{figure}

Integrated over time, both demic and diffusive processes are equally
relevant for most regions. No region, however, shows a demic
contribution of less than 30\%, and all regions have at least a cultural
contribution of more than 15\%. 90\% of all regions show no dominance of
either demic or cultural diffusion (\hyperref[fig:demiccultural]{Figure
3}). A dominance of demic diffusion is evident in the Sahara, and the
Hindukush and other regions around the Central Asian Plateau. Cultural
diffusion is persistent on the Arabian pensinsual, South and Southeast
Asia, and a several regions in southern Siberia and north of the Aral
Sea.

\section{Discussion}\label{discussion}

During each regional transition, both cultural and demic processes play
a role, often even contribute sequentially to a regional agropastoral
transition. In only very few regions, the simulated transition is best
explained by either demic or cultural diffusion processes. Previous
attempts to prove either demic or diffusion processes as solely
responsible for regional agropastoral conditions seem too short-fetched,
when the spatial and temporal interference of cultural and diffusive
processes might have left a complex imprint on the genetic, linguistic
and artifactual record.

In this respect, we confirm Ammerman and Cavalli-Sforza (1973)'s
suggestion and Fort (2012)'s analysis of a probably mixed process
underlying the expansion of agropastoralism and herding. The new finding
here is that for most regions within Eurasia both processes are active,
often contemporaneously, or subsequently, and that a time integrated
view (such as population genetic or linguistic analyses) only picks out
the few regions where either process dominates. For most regions,
however, all of the complex interplay between cultural and demic
diffusion is hidden in a time-integrated view.

This time-integrated view is, however, the only information that is
accessible from radiocarbon arrival date compilations and most model
simulations. Fort (2015), e.g., analysed the variations in diffusion
speeds and attributed these to predominant cultural, demic, or mixed
diffusion for slow, intermediate, and fast apparent diffusion rates,
respectively (Fort 2015). Theirs and our analysis indicate potentially
more demic exchange within Iberia and Northern Italy separated by
predominant cultural or mixed exchange in Southern France; at the coarse
scale of the model regions, however, this comparison should not be
expected to yield conclusive insights.

Based on this time-integrated view, ancient DNA work (e.g. Bramanti et
al. 2009), infers a demic signal throughout Europe. As time control is
difficult in this record, the demic signal might have occured before the
expansion of agropastoralists by migrations of Mesolithic
hunter-gatherers or horticulturalists, or even later. The Y-chromosomal
and the mitochondrial DNA data show different expansion patterns and can
be attributed to multiple migration events, including pre-Neolithic and
post-Neolithic demic events (Sz{é}cs{é}nyi-Nagy et al. 2014), although
most of the introduced variability in the European gene pool was well
established by the Bronze Age (Ricaut 2012).

Migration might have to be functionally disconnected from the spread of
agropastoralism (Gronenborn 2011). Our simulations show that it is not
necessarily only one migration wave and another cultural diffusion event
that shaped the expansion of agropastoralism, but a multitude of
combined events, sometimes more demic, some times more culturally
dominated. This two-faceted expansion process then explains both
archaeogenetic data as well as cultural diffusion evidence, without
requiring distinct migratory processes before the expansion of
agropastoralism.

In GLUES, I did not consider maritime migration, because the Iberian
arrival dates could largely be reconstructed without explicitly
including this process in the model because a secondary wave of advance
enters Iberia from Gibraltar (there are artificial land bridges
connecting across the strait of Gibraltar, Bosporus and the English
Channel to compensate for the lack of maritime transport), which
possibly emulates the fast leap-frog maritime that has been proposed for
that region {[}Battalia2008{]}. For the purpose of investigating
intracontinental diffusion processes in a compact land mass like
Eurasia, the addition of a coastal or sea-mediated additional spread is
not required.

The diachronic view of exchange processes presented here may help to
identify individual migration and cultural exchange processes better
than a time-integrated view. Thus, evidence of trade and exchange
between two cultural layers with genetic continuity does not necessarily
exclude demic diffusion during the entire period of interest, nor does a
different genetic signal imply that cultural diffusion did not take
place, or did not take place at other times.

Where do we see preferential cultural or demic diffusion in this study?
Very roughly, mountaineous regions seem to favour demic diffusion in the
model simulation when integrated over time
(\hyperref[fig:demiccultural]{Figure 3}). This is especially visible for
the Central Asian plateau and its ridges. The Alps, the Pyrenees, the
Iranian Plateau fit this pattern. Other important mountain regions, such
as Anatolia or the Indian Ghats do not exhibit preferential demic
diffusion.

Together with the apparent preferred demic diffusion is the western
Sahara this possibly gives a hint that a lack of local adoption (due to
environmental contraints) could be possible reason for slower or lesser
cultural diffusion. This does not explain, however, the preferential
cultural diffusion in the (also environmentally marginal) Arabian
peninsula. Clearly, more work both in situ and in silico has to be done
to explore the possibility of an environmental constraint selecting for
a specific diffusion process.

These simulations have been performed without being confronted with
sufficient regional archaeological data for most parts of Eurasia, and
the parameters values have been tuned to best reproducing the origin
locations and times of agropastoralism. Only European radiocarbon dates
were used to estimate the diffusion coefficients for the demic, cultural
and mixed diffusion scenarios (see appendix). One Eurasian region tested
for model skill is the Indus region (Lemmen and Khan 2012), and there
the model appears slightly too fast compared to the (often very
uncertain) dates; In a non-Eurasian study (Lemmen 2013) found that
radiocarbon dates for the transitory period 1000~BC--AD 1000 in Eastern
North America were successfully simulated, again with a small model bias
towards earlier dates.

The overall simulation for Eurasia is thus \emph{realistic} in the sense
of providing a consistent spatio-temporal view of one expectation of
prehistoric developments (from a Eurocentric view) at a large scale. The
results are not \emph{real} in the sense that they provide the exact
historical trajectory that has been found at the local scale (cmp
Ackland et al. 2007). The great challenge and promise arising from the
simulation is thus to confront the expectation from the model with the
realization in the archaeological record: only when both disagree can we
learn that either the model is not performing well enough, or that there
is a process that is emancipated from the environmental and cultural
context: then we have quantified human agency. The individual or
society-level decision to migrate or to communicate should be expected
to be at least as rich and complex as the cultural-demic diffusion
picture appearing from a simulation.

\section{Conclusion}\label{conclusion}

I presented a numerical simulation study on the diffusion processes
during the Neolithization in Eurasia, using an adaptive model of
prehistoric societies in their environmental context that is able to
resolve local innovation, cultural diffusion and demic diffusion.
Although a mixed diffusion process had been suggested already long ago,
the analysis of simulations with either cultural or demic diffusion, and
with mixed diffusion, reveals an even more complex spatio-temporal
pattern of the expansion of agropastoralism throughout Eurasia than has
previously been found: demic and cultural processes occur
contemporaneously, or multiple times iteratively or intermittently in
most regions of Eurasia. There is no simple demic or cultural
explanation, but a very complex and rich interplay of both processes in
time and space. The polarized debate of either demic or cultural
diffusion should give way to acknowledging again this more complex
picture and to study and appreciate the richness of mechanisms.

\section{Appendix}\label{appendix}

The diffusion process between a region $i$ another region in its
neighbourhood $j \in \mathcal{N}$ is realized with three diffusion
equations, representing communication, trade, and migration. Diffusion
depends on the influence difference (Renfrew and Level 1979), where
influence is defined as the product of population density $P$ and
technology $T$. The diffusion flux $f$ is proportional to the influence
difference relative to the average influence of regions $i,j$ times
geographically determined conductance between the two regions.

The entries for $c_{i,j}$ in the conductance matrix $\mathcal{C}$
between two regions $i,j$ are constructed from the common boundary
length $L_{i,j}$ divided by the mean area of the regions
$\sqrt{A_i A_j}$. As in Etten and Hijmans (2010), geographically not
connected regions have zero conductance; to connect across the Strait of
Gibraltar, the English Channel, and the Bosporus, the respective entries
in $\mathcal{C}$ were calculated as if a narrow land bridge connected
them.

No additional account is made for increased connectivity along rivers
(Davison et al. (2006), Silva and Steele (2014)), as the regional setup
of the model is biased (through the use of net primary productivity
(NPP) similarity clusters) toward elongating regions in the diretion of
rivers. Altitude and latitude effects are likewise implicitly accounted
for by the NPP clustering in the region generation.

Finally, if the flux between $i,j$ is negative, it is directed inward
from $j$ to $i$, else outward from $i$ to $j$.

\begin{equation}
f_{i,j} =  c_{i,j} \left(\displaystyle{\frac{(P_i T_i A_i + P_j T_j A_j)}{A_i + A_j}}- P_j T_j\right).
\end{equation}

\textbf{Trade/information exchange}: Trait value differences in all
traits $X$ between $i$ and all its neighbours $j$ are summed and added
to region $i$'s trait value.

\begin{equation}
\left.\displaystyle{\frac{\mathrm{d}X_{i}}{\mathrm{d}t}}\right\vert_{\mathrm{trade}} =
\sigma_\mathrm{trade} \sum_{j\in\mathcal{N}_{i},f_{ij}>0} f_{ij}\cdot(X_{j}-X_{i})
\end{equation}

\noindent The parameter $\sigma_{\mathrm{trade}}$ needs to be estimated
(see below); trade is not mass-conserving.

\textbf{Migration} is composed of immigration or emigration, depending
on the sign of the diffusion flux $f$.

\begin{equation}
\left.\displaystyle{\frac{\mathrm{d}P_{i}}{\mathrm{d}t}}\right\vert_{\mathrm{demic}}
=\sigma_\mathrm{demic} \sum_{j\in\mathcal{N}_{i},f_{ij}>0} f_{ij}P_{j}\frac{A_{j}}{A_{i}}
- \sum_{j\in\mathcal{N}_{i},f_{ij}<0} f_{ij}P_{i}.
\end{equation}

\noindent The free parameter $\sigma_{\mathrm{demic}}$ can be chosen to
adjust the speed of migration (see below). Population is redistributed
by scaling with region area $A$, thus, migration is mass-conserving.

\textbf{Hitchhiking traits}: Whenever people move in a demic process,
they carry along their traits to the receiving region:

\begin{equation}
\left.\displaystyle{\frac{\mathrm{d}X_{i}}{\mathrm{d}t}}\right\vert_{\mathrm{demic}} =
\sigma_\mathrm{demic} \sum_{j\in\mathcal{N}_{i},f_{ij}>0} f_{ij}X_{j}\frac{P_{j}A_{j}}{P_{i}A_{i}}
\end{equation}

\subsection{Spread parameter
estimation}\label{spread-parameter-estimation}

Suitable values for the spread parameters are assessed after all other
model parameters have been fixed (for the equations and parameters not
directly relevant to the demic/diffusive analysis, see the supporting
online material provided as a supplement to Lemmen, Gronenborn, and
Wirtz (2011)).

We initially assume that information travels two orders of magnitude
faster than people, based on the typical size of exchange networks (1000
km, Mauvilly, Jeunesse, and Doppler (2008), Gronenborn (1999)), the
average active life time of a tradesperson (order 10 years), and the
comparison with the typical demic front speed of the order 1 km per year
(Ammerman and Cavalli-Sforza (1973)). Starting with this fixed relation
between $\sigma_\mathrm{trade}$ and $\sigma_\mathrm{demic}$, we vary
both parameters such the we get the highest correlation with the dataset
by Pinhasi, Fort, and Ammerman (2005) on European sites; with
$\sigma_\mathrm{trade}=0.2$ and $\sigma_\mathrm{trade}=0.002$ the
highest correlation achieved is $r^2=0.61$ ($n=631$, $p<0.01$). Analysis
of the simulation confirms that this is a parameterisation that
describes \textbf{mixed diffusion} (Lemmen, Gronenborn, and Wirtz
(2011), their figure 6).

For a purely \textbf{demic diffusion} model, trade was switched off
($\sigma_\mathrm{trade}=0$) and $\sigma_\mathrm{demic}$ was varied
(systematically increased) to again obtain the best correlation with the
data. The estimated parameter value is $\sigma_\mathrm{demic}=0.008$.
The respective procedure was applied to estimate the parameter
$\sigma_\mathrm{trade}$ for a purely \textbf{cultural diffusion}
best-fitting model; its value was determined to be $0.3$.

\section{References}\label{references}

\setlength{\parindent}{-0.2in} \setlength{\leftskip}{0.2in}
\setlength{\parskip}{8pt} \vspace*{-0.2in} \noindent

Ackland, Graeme J, Markus Signitzer, Kevin Stratford, and Morrel H
Cohen. 2007. ``Cultural Hitchhiking on the Wave of Advance of Beneficial
Technologies.'' \emph{Proc. Natl. Acad. Sci. U. S. A.} 104(21):
8714--19.

Ammerman, A J, and Luigi Luca Cavalli-Sforza. 1973. ``A Population Model
for the Diffusion of Early Farming in Europe.'' In \emph{Explan. Cult.
Chang.}, ed. C Renfrew. London: Duckworth, 343--57.

Barker, Graeme. 2006. \emph{The Agricultural Revolution in Prehistory:
Why Did Foragers Become Farmers?} Oxford, United Kingdom: Oxford
University Press.

Battaglia, Vincenza, Simona Fornario, Nadia Al-Zahery, Anna Olivieri,
Peter A Underhill, and Ornella Semino. 2008. ``Y-Chromosomal Evidence of
the Cultural Diffusion of Agriculture in Southeast Europe.'' \emph{Eur.
J. Hum. Genet.} (November): 1--11.

Bocquet-Appel, Jean-Pierre, Stephan Naji, Marc {Vander Linden}, and
Janusz Kozlowski. 2012. ``Understanding the Rates of Expansion of the
Farming System in Europe.'' \emph{J. Archaeol. Sci.} 39(2): 531--46.

Bramanti, Barbara, M G Thomas, Wolfgang Haak, M Unterlaender, P Jores, K
Tambets, I Antanaitis-Jacobs, M N Haidle, R Jankauskas, C-J Kind, F
Lueth, T Terberger, J Hiller, S Matsumura, P Forster, and J Burger.
2009. ``Genetic Discontinuity Between Local Hunter-Gatherers and Central
Europe's First Farmers.'' \emph{Science (80-. ).} 326(5949): 137--40.

Brovkin, Victor, Andrei Ganopolski, and Yuri Svirezhev. 1997. ``A
Continuous Climate-Vegetation Classification for Use in
Climate-Biosphere Studies.'' \emph{Ecol. Modell.} 101: 251--61.

Chikhi, Lounes, Richard A Nichols, Guido Barbujani, and Mark A Beaumont.
2002. ``Y Genetic Data Support the Neolithic Demic Diffusion Model.''
\emph{Proc. Natl. Acad. Sci.} 99(17): 11008--13.

Childe, Vere Gordon. 1925. \emph{Dawn Of European Civilization}. 1st ed.
Routledge {[}reprinted 2005{]}.

Clark, J G D. 1965. ``Radiocarbon Dating and the Expansion of Farming
Culture from the Near East over Europe.'' \emph{Proc. Prehist. Soc.} 31:
57--73.

Davison, Kate, Pavel M Dolukhanov, and GR Sarson. 2009. ``Multiple
Sources of the European Neolithic: Mathematical Modelling Constrained by
Radiocarbon Dates.'' \emph{Quat. Int.} 44: 1--17.

Davison, Kate, Pavel M Dolukhanov, Graeme R. Sarson, and Anvar Shukurov.
2006. ``The Role of Waterways in the Spread of the Neolithic.'' \emph{J.
Archaeol. Sci.} 33(5): 641--52.

Deguilloux, Marie-France, Rachael Leahy, Marie-H{é}l{è}ne Pemonge, and
St{é}phane Rottier. 2012. ``European Neolithization and Ancient DNA: An
Assessment.'' \emph{Evol. Anthropol.} 21(1): 24--37.

Edmonson, Munro S. 1961. ``Neolithic Diffusion Rates.'' \emph{Curr.
Anthropol.} 2(2): 71--102.

Etten, Jacob van, and Robert J. Hijmans. 2010. ``A Geospatial Modelling
Approach Integrating Archaeobotany and Genetics to Trace the Origin and
Dispersal of Domesticated Plants.'' \emph{PLoS One} 5(8): e12060.

Fort, Joaquim. 2012. ``Synthesis Between Demic and Cultural Diffusion in
the Neolithic Transition in Europe.'' \emph{Proc. Natl. Acad. Sci. U. S.
A.} 109(46): 18669--73.

---------. 2015. ``Demic and Cultural Diffusion Propagated the Neolithic
Transition Across Different Regions of Europe.''

Fu, Qiaomei, Pavao Rudan, Svante P{ä}{ä}bo, and Johannes Krause. 2012.
``Complete Mitochondrial Genomes Reveal Neolithic Expansion into
Europe.'' \emph{PLoS One} 7(3): e32473.

Fuller, Dorian Q, Tim Denham, Manuel Arroyo-kalin, Leilani Lucas, Chris
J Stevens, Ling Qin, and Robin G Allaby. 2014. ``Convergent Evolution
and Parallelism in Plant Domestication Revealed by an Expanding
Archaeological Record.'' 111(17).

Galeta, Patrik, Vladim{í}r Sl{á}dek, Daniel Sosna, and Jaroslav Bruzek.
2011. ``Modeling Neolithic Dispersal in Central Europe: Demographic
Implications.'' \emph{Am. J. Phys. Anthropol.} 146(1): 104--15.

Gronenborn, Detlef. 1999. ``A Variation on a Basic Theme: The Transition
to Farming in Southern Central Europe.'' \emph{J. World Prehistory}
13(2): 123--210.

---------. 2011. ``Neolithic Western Temperate and Central Europe.'' In
\emph{Oxford Handb. Archae- Ology Anthropol. Hunt. Gatherers}, ed. M.
{Cummings, V., Jordan, P., Zvelebil}. Oxford University Press.

Haak, Wolfgang, Oleg Balanovsky, Juan J. Sanchez, Sergey Koshel, Valery
Zaporozhchenko, Christina J. Adler, Clio S. I. {Der Sarkissian}, Guido
Brandt, Carolin Schwarz, Nicole Nicklisch, Veit Dresely, Barbara
Fritsch, Elena Balanovska, Richard Villems, Harald Meller, Kurt W. Alt,
and Alan Cooper. 2010. ``Ancient DNA from European Early Neolithic
Farmers Reveals Their Near Eastern Affinities'' ed. David Penny.
\emph{PLoS Biol.} 8(11): e1000536.

Hodder, Ian. 1990. \emph{The Domestication of Europe. Structure and
Contingency in Neolithic Societies}. Oxford: Blackwell.

Lemmen, Carsten. 2013. ``Mechanisms Shaping the Transition to Farming in
Europe and the North American Woodland.'' \emph{Archaeol. Ethnol.
Anthropol. Eurasia} 41(3): 48--58.

Lemmen, Carsten, and Aurangzeb Khan. 2012. ``A Simulation of the
Neolithic Transition in the Indus Valley.'' In \emph{Clim. Landscapes,
Civilizations}, Geophysical Monograph Series, eds. Liviu Giosan, Dorian
Q Fuller, Kathleen Nicoll, Rowan K Flad, and Peter D Clift. Washington:
American Geophysical Union, 107--14.

Lemmen, Carsten, Detlef Gronenborn, and Kai W Wirtz. 2011. ``A
Simulation of the Neolithic Transition in Western Eurasia.'' \emph{J.
Archaeol. Sci.} 38(12): 3459--70.

Mauvilly, Michel, Christian Jeunesse, and Thomas Doppler. 2008. ``Ein
Tonstempel Aus Der Spätmesolithischen Fundstelle von Arconciel / La
Souche (Kanton Freiburg, Schweiz ).'' \emph{Quartär} 55: 151--57.

New, M, MC Todd, M Hulme, and P Jones. 2001. ``Precipitation
Measurements and Trends in the Twentieth Century.'' \emph{Int. J.
Climatol.} 21: 1899--1922.

Patterson, M. A., Graeme R. Sarson, H.C. Sarson, and Anvar Shukurov.
2010. ``Modelling the Neolithic Transition in a Heterogeneous
Environment.'' \emph{J. Archaeol. Sci.} 37(11): 2929--37.

Pinhasi, Ron, Joaquim Fort, and A J Ammerman. 2005. ``Tracing the Origin
and Spread of Agriculture in Europe.'' \emph{Public Libr. Sci. Biol.}
3(12): e410.

Renfrew, Colin, and E V Level. 1979. ``Exploring Dominance: Predicting
Polities from Centers.'' In \emph{Transform. Math. Approaches to Cult.
Chang.}, eds. Colin Renfrew and K Cooke. New York: Academic Press,
145--66.

Renfrew, Colin, ed. 1987. ``Archaeology and Language. The Puzzle of
Indo-European Origins.'': 346.

Ricaut, Fran{ç}ois-Xavier. 2012. ``A Time Series of Prehistoric
Mitochondrial DNA Reveals Western European Genetic Diversity Was Largely
Established by the Bronze Age.'' \emph{Adv. Anthropol.} 02(01): 14--23.

Rowley-Conwy, Peter. 2004. ``How the West Was Lost.'' \emph{Curr.
Anthropol.} 45(October): 83--113.

Silva, Fabio, and James Steele. 2014. ``New Methods for Reconstructing
Geographical Effects on Dispersal Rates and Routes from Large-Scale
Radiocarbon Databases.'' \emph{J. Archaeol. Sci.} 52: 609--20.

Sz{é}cs{é}nyi-Nagy, Anna, Guido Brandt, Victoria Keerl, J{á}nos Jakucs,
Wolfgang Haak, Marc Fecher, Sabine Moeller-Rieker, Kitti K{ö}hler,
Bal{á}zs Guszt{á}v Mende, Kriszti{á}n Oross, Tibor Marton, Anett
Oszt{á}s, Vikt{ó}ria Kiss, Gy{ö}rgy P{á}lfy, Erika Moln{á}r, P{á}l
Raczky, Alexandra Anders, Katalin Sebők, Andr{á}s Czene, Roz{á}lia
Kust{á}r, Tibor Paluch, Krisztina Somogyi, Mario {Š}laus, Martin Novak,
Zsuzsanna Zoffmann, G{á}bor T{ó}th, Brigitta Ősz, Vanda Voicsek, Eszter
B{á}nffy, and Kurt W. Alt. 2014. ``Tracing the Genetic Orgigin of
Europe's First Farmers Reveals Insights into Their Social Structure.''

Wirtz, Kai W, and Bruno Eckhardt. 1996. ``Effective Variables in
Ecosystem Models with an Application to Phytoplankton Succession.''
\emph{Ecol. Modell.} 92(1): 33--53.

Wirtz, Kai W, and Carsten Lemmen. 2003. ``A Global Dynamic Model for the
Neolithic Transition.'' \emph{Clim. Change} 59(3): 333--67.

Zvelebil, Marek. 1998. ``What's in a Name: The Mesolithic, the
Neolithic, and Social Change at the Mesolithic-Neolithic Transition.''
In \emph{Underst. Neolit. North-Western Eur.}, ed. C {Edmonds, M,
Richards}. Glasgow: Cruithne Press, 1--36.

\ifreview\else\end{twocolumn}\fi
\end{document}